
\magnification=\magstep1
\headline={\ifnum\pageno=1\hfil\else\hfil\tenrm--\ \folio\ --\hfil\fi}
\footline={\hfil}
\hsize=6.0truein
\vsize=8.54truein
\hoffset=0.25truein
\voffset=0.25truein
\baselineskip=20pt
%
%
\tolerance=9000
\hyphenpenalty=10000
%
%
%
\font\eightrm=cmr8
\font\mbf=cmmib10 \font\mbfs=cmmib10 scaled 833
\font\msybf=cmbsy10 \font\msybfs=cmbsy10 scaled 833
\font\smcap=cmcsc10
%
%
%
\def\bmit{\fam9}
\textfont9=\mbf \scriptfont9=\mbfs \scriptscriptfont9=\mbfs
\def\bmsy{\fam10}
\textfont10=\msybf \scriptfont10=\msybfs \scriptscriptfont10=\msybfs
%
%
\mathchardef\alpha="710B
\mathchardef\beta="710C
\mathchardef\gamma="710D
\mathchardef\delta="710E
\mathchardef\epsilon="710F
\mathchardef\zeta="7110
\mathchardef\eta="7111
\mathchardef\theta="7112
\mathchardef\iota="7113
\mathchardef\kappa="7114
\mathchardef\lambda="7115
\mathchardef\mu="7116
\mathchardef\nu="7117
\mathchardef\xi="7118
\mathchardef\pi="7119
\mathchardef\rho="711A
\mathchardef\sigma="711B
\mathchardef\tau="711C
\mathchardef\upsilon="711D
\mathchardef\phi="711E
\mathchardef\chi="711F
\mathchardef\psi="7120
\mathchardef\omega="7121
\mathchardef\varepsilon="7122
\mathchardef\vartheta="7123
\mathchardef\varpi="7124
\mathchardef\varrho="7125
\mathchardef\varsigma="7126
\mathchardef\varphi="7127
\mathchardef\nabla="7272
\mathchardef\cdot="7201
\def\grad{\nabla} 
%
%
%
\def\ldb{[\![}
\def\rdb{]\!]}
%
%
\def\bigldb{\bigl[\!\bigl[}
\def\bigrdb{\bigr]\!\bigr]}
%
%
\def\biggldb{\biggl[\!\!\biggl[}
\def\biggrdb{\biggr]\!\!\biggr]}
\def\um{{\,\mu\rm m}}
\def\cm{{\rm\,cm}}
\def\km{{\rm\,km}}
\def\au{{\rm\,AU}}
\def\pc{{\rm\,pc}}
\def\kpc{{\rm\,kpc}}
\def\mpc{{\rm\,Mpc}}
\def\sec{{\rm\,s}}
\def\yr{{\rm\,yr}}
\def\gm{{\rm\,g}}
\def\kms{{\rm\,km\,s^{-1}}}
\def\kelvin{{\rm\,K}}
\def\erg{{\rm\,erg}}
\def\ev{{\rm\,eV}}
\def\hz{{\rm\,Hz}}
\def\msun{{\,M_\odot}}
\def\lsun{{\,L_\odot}}
\def\rsun{{\,R_\odot}}
%
%
\def\degree{^\circ}
\def\arcsec{''\hskip-3pt .}
%
%
\newcount\eqnumber
\eqnumber=1
%
\def\new{{\the\eqnumber}\global\advance\eqnumber by 1}
%
%
\def\ref#1{\advance\eqnumber by -#1 \the\eqnumber
     \advance\eqnumber by #1 }
%
%
\def\last{\advance\eqnumber by -1 {\the\eqnumber}\advance
     \eqnumber by 1}
%
%
\def\eqnam#1{\xdef#1{\the\eqnumber}}
%
%
%
\def\refindent{\par\noindent\hangindent=3pc\hangafter=1 }
\def\aa#1#2#3{\refindent#1, A\&A, #2, #3}
\def\aasup#1#2#3{\refindent#1, A\&AS, #2, #3}
\def\aj#1#2#3{\refindent#1, AJ, #2, #3}
\def\apj#1#2#3{\refindent#1, {\it Ap. J.}, {\bf#2}, #3.}
\def\apjlett#1#2#3{\refindent#1, {\it Ap. J. (Letters)}, {\bf #2}, #3.}
\def\apjsup#1#2#3{\refindent#1, ApJS, #2, #3}
\def\araa#1#2#3{\refindent#1, ARA\&A, #2, #3}
\def\baas#1#2#3{\refindent#1, BAAS, #2, #3}
\def\icarus#1#2#3{\refindent#1, Icarus, #2, #3}
\def\mnras#1#2#3{\refindent#1, {\it M.N.R.A.S.}, {\bf#2}, #3.}
\def\nature#1#2#3{\refindent#1, {\it Nature}, {\bf #2}, #3.}
\def\pasj#1#2#3{\refindent#1, PASJ, #2, #3}
\def\pasp#1#2#3{\refindent#1, PASP, #2, #3}
\def\qjras#1#2#3{\refindent#1, QJRAS, #2, #3}
\def\science#1#2#3{\refindent#1, Science, #2, #3}
\def\sov#1#2#3{\refindent#1, Soviet Astr., #2, #3}
\def\sovlett#1#2#3{\refindent#1, Soviet Astr.\ Lett., #2, #3}
\def\refpaper#1#2#3#4{\refindent#1, #2, #3, #4}
\def\refbook#1{\refindent#1}
%
\def\sect#1 {
  \bigbreak
  \centerline{\bf #1}
  \bigskip}
\def\subsec#1#2 {
  \bigbreak
  \centerline{#1.~{\it #2}}
  \bigskip}
\newcount\figno
\figno=0
\def\figure{\global\advance\figno by 1 Figure~\the\figno.~}
%
%
\def\degs{$^\circ$}
\def\>{$>$}
\def\<{$<$}
\def\bsl{$\backslash$}
\def\simlt{\lower.5ex\hbox{$\; \buildrel < \over \sim \;$}}
\def\simgt{\lower.5ex\hbox{$\; \buildrel > \over \sim \;$}}
\def\sqr#1#2{{\vcenter{\hrule height.#2pt
      \hbox{\vrule width.#2pt height#1pt \kern#1pt
         \vrule width.#2pt}
      \hrule height.#2pt}}}
\def\square{\mathchoice\sqr34\sqr34\sqr{2.1}3\sqr{1.5}3}
%
%
\def\vek#1{{\bfsl #1}}
\def\today{\ifcase\month\or
	January\or February\or March\or April\or May\or June\or
	July\or August\or Setrueptember\or October\or November\or December\fi
	\space\number\day, \number\year}
%
\def\head#1{\headline={\ifnum\pageno>1
	{\tenrm #1} \hfil Page \folio
	\else\hfil\fi}}
%
\def\ref #1;#2;#3;#4{\par\pp #1, {\it #2}, {\bf #3}, #4}
\def\book #1;#2;#3{\par\pp #1, {\it #2}, #3}
\def\rep #1;#2;#3{\par\pp #1, #2, #3}
%
%
\def\undertext#1{$\underline{\smash{\hbox{#1}}}$}
\def\eol{\hfil\break}
\def\EFtoday{EF\number\month/\number\day/\number\year}
\newbox\grsign \setbox\grsign=\hbox{$>$}
\newdimen\grdimen \grdimen=\ht\grsign
\newbox\laxbox \newbox\gaxbox
\setbox\gaxbox=\hbox{\raise.5ex\hbox{$>$}\llap
     {\lower.5ex\hbox{$\sim$}}}\ht1=\grdimen\dp1=0pt
\setbox\laxbox=\hbox{\raise.5ex\hbox{$<$}\llap
     {\lower.5ex\hbox{$\sim$}}}\ht2=\grdimen\dp2=0pt
\def\gax{\mathrel{\copy\gaxbox}}
\def\lax{\mathrel{\copy\laxbox}}
\def\etal{{\it et~al.\ }}

\def\ppcc{\sigma_{\gamma\gamma}}
\def\gamfac{\gamma(1-\beta \cos\theta)}
\def\gamfacw{\gamma(1-\beta \cos\omega)}
\def\gamfacp{\gamma(1-\beta \cos\psi)}
\def\logNP{$\log N$--$\log P$}
\def\xellc{R_o}
%

%
\centerline{\bf The X-ray Spectrum of the Soft Gamma Repeater 1806-20}
\vskip 0.5cm
\centerline{E. E. Fenimore, J. G. Laros }
\centerline{Los Alamos National Laboratory, MS D436, Los Alamos, NM 87545}
\vskip 0.5cm
\centerline {and}
\vskip 0.5cm
\centerline{A. Ulmer}
\centerline{Princeton University, Princeton NJ 08544}
\vskip 0.5cm

\centerline{{\it in press} Astrophysical Journal (9/94)}
\centerline{(tables will appear in published version only)}
\vskip 0.5cm

\centerline{\bf ABSTRACT}
\vskip 0.5cm

Soft Gamma Repeaters (SGRs) are a class of rare, high-energy galactic
transients that have episodes of short ($\sim 0.1$ sec), soft ($\sim 30$ keV),
intense ($\sim 100$ Crab), gamma-ray bursts.
We report an analysis of the x-ray emission from 95 SGR1806-20 events
observed by the {\it International Cometary Explorer}.
The spectral shape remains remarkably constant for bursts that differ
in intensity by a range of 50.
Below $\sim$ 15 keV the number spectrum falls off rapidly such that we
can estimate the total intensity of the events.
Assuming that SGR1806-20 is associated with the supernova remnant G10.0-0.3
(Kulkarni and Frail, Murakami \etal), the brightest events had a total
luminosity of
$1.8 \times 10^{42}$ erg sec$^{-1}$, a factor of $2 \times 10^{4}$ above the
Eddington limit.
A third of the emission was above 30 keV.
There are at least three processes that are consistent with the spectral
rollover below 15 keV.
(1)~The rollover is consistent with some forms of self absorption.
Typical thermal temperatures are $\sim$ 20 keV and require an emitting
surface with a radius
between 10 and 50 km.
The lack of spectral variability implies that only the size of the
emitting surface varies from event to event.
If the process is thermal synchrotron the required magnetic field might be
too small to confine the plasma against the super Eddington flux.
(2)~The low energy rollover could be due to photoelectric absorption by
$10^{24}$ Hydrogen atoms cm$^{-2}$ of neutral material with a cosmic abundance.
This assumes a
continuum similar to thermal bremsstrahlung with a temperature of
$\sim 22$ keV.
The material is most likely to be associated with the object as circumstellar
matter a few A.U.~from the central source rather than
foreground clouds or directly at the site of the energy
release.
(3)~Emission in the two lowest harmonics from a $1.3 \times 10^{12}$ Gauss
field
would appear as Doppler broadened lines and fall off rapidly below 15 keV.

\medskip
{\it Subject Headings:} Gamma-Rays: Bursts -- X-Rays: Bursts -- stars:
individual (SGR1806-20)
\vfill
\eject

\sect{1.~INTRODUCTION}

Mazets and Golenetskii (1981) first suggested that
short events comprise a separate class
of gamma ray bursts (GRBs) based on soft events from
GRB790305 (the ``March fifth'' event),
GRB790107, and GRB790324 as well as several hard events.
Two of these sources had soft repetitions: SGR790305
(Mazets and Golenetskii 1981; Golenetskii, Iiyinskii,
and Mazets 1984) and GRB790324 (Mazets, Golenetskii, and Guryan 1981).
With the discovery of $\sim$ 110 soft and short recurrences
from SGR790107
(Laros \etal 1986, 1987; Atteia \etal 1987),
this class has been called the Soft Gamma Repeaters
(SGR).
Norris \etal 1991 has reviewed their properties.
The sources are designated SGR followed by the location.
Thus, GRB790305 (March fifth) is also known as SGR0526-22, GRB790107 is
known as SGR1806-20, and GRB790324 is SGR1900+14.
These sources are thought to form a separate class of events
with different physics than the classic GRB.
Other short GRBs (i.e.,
less than 1 second in duration) have hard spectra like the longer classic GRBs
and are probably just
extreme examples of a bimodal duration distribution for classic GRBs
(Klebesadel 1992, Kouveliotou \etal 1993b).
Recently, the Burst and Transient Experiment (BATSE) has observed some
activity in both SGR1900+14 and SGR1806-20 (Kouveliotou \etal 1993a,
Kouveliotou \etal 1994).

The Soft Gamma Repeaters have distinct
properties from those of other
classes of high energy transients.
Their typical photon energy is 30 keV whereas the x-ray bursters have
typical
energies of 3 keV and the classical GRBs have typical energies in excess
of 300 keV (see {\it e.g.,} Band \etal 1993).
The classical GRBs often have complex time histories lasting from less than
a second to more than 1000 seconds (Klebesadel, 1992, Fishman \etal 1993)
and the x-ray bursters have simpler
time histories with a sharp rise and a decaying tail lasting about 30
seconds.
In contrast, time histories of SGR rise
sharply and fall with the whole event lasting on the order of
0.1 second (Atteia \etal 1987). Classic GRBs do not seem to repeat whereas
x-ray bursters do
repeat with a characteristic pattern (Lewin and Joss, 1983).
The SGR do not have any discernible pattern in the
repetitions (Laros \etal 1987).

The bright March fifth event has been studied in great detail.
A number of interesting
characteristics were discovered that have helped
guide the study of soft gamma repeaters.
Foremost, there is an 8 second periodic emission for more than
200 seconds after the initial burst (Mazets \etal 1979).
If the pulsations are due to stellar rotation, then such a period strongly
suggests a neutron star origin.
The March fifth error box is extremely small (0.1 arcmin$^{2}$)
(Cline, \etal 1982) and occurs
in the direction of the supernova remnant N49 in the Large Magellanic
Cloud (Evans \etal 1980). The distance to N49 (55 kpc) implies
a huge energy release in the initial
spike ($\sim 10^{44}$ ergs sec$^{-1}$ above 30 keV).
Such a release of energy on a neutron star would be super-Eddington by a
factor of $10^{6}$, raising doubts that the object could be as far away as
implied by the supernova remnant in its error box.

SGR1806-20 is located within 7 degrees of the
galactic center (Atteia \etal 1987) and the third
known
burster, SGR1900+14, is in the galactic plane (Mazets, Golenetskii, and
Guryan 1979)
Therefore, a population I distribution has been suggested
(Laros \etal 1986, Kouveliotou \etal 1987).
The recent detection of a burst in a small error box centered on the
supernova remnant G10.0-0.3 (Murakami \etal 1994) confirms the earlier
association suggested by Kulkarni and Frail (1993).
The radio characteristics of G10.0-0.3
provides a distance estimate of 17 kpc for
SGR1806-20 (Kulkarni and Frail, 1993).

\sect{2.~INSTRUMENTATION}

Fortunately, SGR1806-20 was located in the ecliptic plane, and thus, was always
within the field of view of the UCB/Los Alamos experiment on
the {\it International Cometary Explorer} (ICE, see Anderson \etal 1978).
This experiment
provided nearly continuous
coverage from late 1978 to 1986, and all known bursts from SGR1806-20 were
detected by ICE. The experiment consisted of a proportional counter and a
scintillator which together
gave spectral information between 5 keV and 2 MeV.
The ICE experiment yielded excellent data in which to search for
spectral variability and to determine the overall spectral shape of SGR1806-20.
Laros \etal (1987) reported 111 bursts believed to be
from SGR1806-20. Later, a more sensitive search found 23
more events
in ICE data (Laros \etal 1990), and since the 111 bursts showed a power
law luminosity function, it is likely that more existed below the
sensitivity of the instrument.
Ulmer \etal (1993) cataloged all of the ICE SGR1806-20 events.
We studied a subset of the original 111 bursts.
Some of the events were rejected because of background variations (most
likely due to solar activity)
which made estimating the signals uncertain, and a few events were not
used because the instrument had been commanded to a nonstandard gain. The
data for a few events are currently unavailable.
In all, 95 of the original
111 were judged suitable for spectral analysis. These bursts
had backgrounds that could
be fit to within the statistics with either a constant or linear function.
The uncertainty due to the background fitting was propagated into the
error bars quoted in this paper.

ICE viewed these bursts with both a collimated proportional counter (PC)
which functioned in the 5 to 14 keV range and a collimated
scintillator counter (SC) which allowed measurements from 26 keV to 2 Mev.
The PC had an effective are of $\sim$ 1.5 cm$^{2}$ and six energy channels:
PC1 (5.0
to 6.0 keV), PC2 (6.0 to 7.0 keV), PC3 (7.0 to 8.5 keV), PC4 (8.5 to 10.0 keV),
PC5 (10.0 to
12.0 keV), and PC6 (12.0 to 14.0 keV).
The SC had an effective area of $\sim$ 22 cm$^{2}$ and twelve energy
channels; however,
the bursts from SGR1806-20 are so soft that
they rarely show a net signal beyond the fifth scintillator energy bin (SC5).
The third SC channel (SC3) stopped operating in March of 1983,
before the source became extremely active.
Thus, usually there are 3 or 4 SC channels with net signals: SC1 (from 25.9
to 43.2
keV), SC2 (43.2 to 77.5 keV), SC4 (121 to 154 keV) and SC5 (154 to 236 keV).
The SC6 channel covered from 236 to 320 keV.
In our fitting, we used 11 energy channels, 6 from the PC and 5 from the SC.
Time samples were continuously taken at half second intervals for most of the
energy channels. Two energy channels, PC5 and PC6,
had time samples of four seconds.

\sect{3.~SPECTRAL VARIABILITY}

A key characteristic of SGRs is the similarity between different bursts from
the same source.
Differences among the
spectra of various bursts from SGR1806-20 are hard to discern.
The bright bursts from SGR1806-20 observed by {\it Prognoz 9}
appear to have a common spectral shape above 30 keV (Atteia \etal 1987).
The spectra from five different {\it
Prognoz 9}
bursts were consistent with optically thin thermal bremsstrahlung
with a single temperature even through
the total intensity of the different bursts in that analysis varied by
a factor of four.
In addition, the {\it Solar Maximum
Mission} (SMM)
resolved one burst from SGR1806-20 into two time samples with
spectral information above 30 keV
and found no indication of spectral evolution
(Kouveliotou \etal 1987).
However, significant variations have been observed within the main peak of
the March 5th event (Fenimore \etal 1981).

These previous studies used the strongest SGR1806-20 events.
The 95 ICE events cover a dynamic range of 50 and allow a more sensitive
search for correlations between brightness and spectral shape.
Figure 1 shows a hardness ratio of counts (SC2 to SC1)
plotted as a function of
estimated fluence for the 95 SGR1806-20 events used in this study.
The fluence is based on the sum of SC1 and SC2
normalized by the integrated spectral shape of the strongest bursts.
Error bars on the estimated
fluence are excluded
from the plot but were used in the analysis.
Fitting a constant to the 95 hardness ratios
yielded a $\chi^{2}$ of 164 with 94 degrees of freedom.
Fitting a linear function
yielded a somewhat better $\chi^2$ (145, 93 degrees of freedom).
Neither fit has an acceptable $\chi^2$.
The solid line in Fig.~1 shows the best fit linear function.
The high $\chi^2$ of the fits seems to be attributed to scatter about a typical
hardness ratio rather than
being due to some nonlinear trend in hardness with fluence.
This scatter could be due to systematic effects.
For example, there are posts within the SC collimator which could
partially intercept the beam from some events depending on the
rotation angle of the satellite at the time of the burst.
Alternatively, the scatter could be due to intrinsic variations within the
source.
If so, this would be the first evidence that the spectra of SGR1806-20 can
vary from
burst to burst.
We want to emphasize that the variation is very small, the linear fit
only varies by 30\% over a dynamic range of 50.
If the changes in fluence were due to
blackbody emission with changing temperature but constant area,
the resulting hardness ratios would vary
by $\sim$ 12.
Clearly, some process is regulating the spectrum and maintaining a constant
shape.
To match the hardness variations  seen in the
linear fit would require the temperature of a thermal
bremsstrahlung continuum to change only from 19 to 28 keV. This is a
small range of spectral shape compared to classic gamma ray bursts which show
variations from 100 to 600 keV and variations within individual
events (Band, \etal 1993).
This relative constancy of the hardness ratios over
such a large range of intensities
is a feature that a successful model must explain.
We have also searched for a correlation between
hardness and
the time of occurrence of the bursts and found nothing obvious.

Another potential correlation would be between the
x-ray observations (with the PC) and the gamma-ray observations (with the SC).
We find that the hardness ratios
between 26 to 78 keV (SC1+SC2) and 10 to 14 keV (PC4+PC5+PC6)
are consistent with a constant spectral shape. We cannot place rigorous
constraints on this ratio because of low statistics in the PC
channels. We would notice a difference if the ratios were correlated between
intensity and varied by more than a factor of 2 over the range of observations.
However, comparing Figs. 5, 6, and 7 shows a trend that the x-ray
emission grows relative to the gamma-ray emission for weaker bursts.

\sect{4.~X-RAY SPECTRUM OF SGR1806-20}

Previous spectral analyses of bursts from SGR1806-20
have been limited to above 30 keV (Mazets, \etal 1982,
Atteia \etal 1987, Kouveliotou \etal 1987)
except for the ICE PC observations of GRB790107 (Laros
\etal 1986).
Laros \etal found that the PC observations showed a deficiency below 30
keV relative to optically thin thermal bremsstrahlung, but the signal was not
strong enough to determine whether or not the spectrum rolled over
and decreased at low energy.
The ICE PC observed SGR1806-20 down to 5 keV.
The x-ray diffuse background was dominant so only the strongest events
could be studied individually.
To search for spectral variability, we ordered the events by the fluence
in SC1 plus SC2 since fluence is usually correlated with signal-to-noise.
We then summed together events to obtain statistically significant spectra.
It was necessary to group together the
strongest event and the third
strongest event. These two events occurred one second apart on
November 16, 1983
and were added together
not because of weak statistics, but because
the PC5 and PC6 temporal samples encompassed both bursts.

We fit different
spectral shapes to the data by folding the models through the instrument
response matrix and varied the model until the best parameters were
found.
It is clear from the brighter events and the summed weaker events that the
number spectrum is no longer monotonically increasing below 15 keV; there
is a strong rollover in the x-ray portion of the spectrum.
All
models without rollovers gave unacceptable $\chi^2$.
This includes thermal bremsstrahlung which had been used exclusively
above 30 keV in previous work as well as
power laws, exponentials, and
unsaturated inverse Comptonization.
We tried many different types of spectral shapes to determine
which physics might be applicable.
The models that produce rollovers can be grouped into three
classes: (1) self absorption (blackbody or various functions connected
to a Rayleigh-Jeans law),
(2) photoelectric absorption by
neutral material in molecular clouds or in a circumstellar region,
and (3) edges to emission processes such as cyclotron radiation.

Figures 2 through 7 show fits to various spectra for SGR1806-20.
The upper right hand corner gives the events that were used.
For example, 831116-G means the 7$^{th}$ event on Nov 11 1983.
The number in parenthesis gives the event's ranking in fluence
({\it i.e.,} 1 means it was the strongest event seen by ICE).
The UT
for these events are given in Table I and
Ulmer \etal (1993) list all the events for each day and their intensities.
In some cases the event was split between two 0.5 sec samples in which case
we only used the sample with the strongest signal to avoid reducing
the signal-to-noise by including a 0.5 sec sample with only
a small fraction of the total signal.
The duration for the purpose of converting counts to photon sec$^{-1}$
was taken to be the duration
of the spectral sample (0.5 sec); their actual durations are listed in Table I.
For these reasons, the area under the curves will not necessarily be the
total intensity of the event as listed in Table I.
Shown in each figure are spectral fits for blackbody (BB),
thermal bremsstrahlung merged with a blackbody (TB-BB), a power law
merged with thermal bremsstrahlung (TB-PL),
and thermal bremsstrahlung with
photoelectric absorption by neutral material with a cosmic abundance (TB-PE).
The vertical scale corresponds to the blackbody fit and
each of the other spectra have
been displaced by a factor of 10 to improve visibility.

The solid curves in the figures are the best fit number spectra
(photons sec$^{-1}$ cm$^{-2}$ keV$^{-1}$).
The data points with error bars are not measurements of the number spectrum
but rather give the residuals between the observed counts and the counts
obtained
from folding the best fit spectrum through the response matrix.
The vertical error bars represent $^+_-1\sigma$ and the distance from the
horizontal central bar to the number spectrum is the residual for that
energy channel in units of $1\sigma$.
The length of the horizontal central bar gives the width of the energy loss
bin.
This representation of the spectrum is ``obliging'' (Fenimore, Klebesadel,
and Laros, 1983, Loredo and Epstein, 1989) in that the data points will
move to agree with whatever spectral shape is assumed
since what they actually represent is the difference between the observed and
the model.
This obliging nature is intrinsic to detectors (such as the PC and SC) that
do not have a $\delta$ function response to monoenergetic photons or have
strongly varying efficiency.

To avoid confusion, SC4, SC5, and SC6 are only shown for the blackbody
fits and only if they are positive.
These points have poor statistics and are particularly obliging.

\subsec{4.1}{Blackbody}

A natural explanation for any spectrum that rolls over is that it is
optically thick self absorption which often approaches a
blackbody spectrum.
A blackbody spectrum also has the key advantage that a distance can be
estimated with only an assumption concerning the size of the emitting region.
A blackbody number spectrum can be expressed as
$$
\phi_{BB}(E) = {{0.008284E^{2}\Psi} \over {e^{E/kT}-1}}    \eqno(\new)
$$
in photons keV$^{-1}$ cm$^{-2}$ sec$^{-1}$ where kT is the
temperature (in keV) and $\Psi$ has units of (km/kpc)$^2$.
Figures 2 though 4 show the blackbody fits for the four strongest events
individually (with the first and third strongest events combined because
PC5 and PC6
were not temporally resolved).
The two blackbody free parameters are the temperature ($\sim 9$ keV) and
$\Psi$ ($\sim 30$ km$^{2}$ kpc$^{-2}$).
In Table II, the errors for the temperature are $^+_-1\sigma$
and are found from where $\chi^2$ changes by 1, appropriate for the
range of a single interesting parameter ({\it cf.} Lampton, Margon,
and Bowyer 1976).  The ``Prob'' in Table II is the probability that the
$\chi^2$ would be as large as observed by random chance, considering the
number of degrees of freedom.
Individually, each burst can be fit with an acceptable $\chi^2$ (see Table II),
although other models with one or two more free parameters have
substantially smaller $\chi^2$.
Furthermore, in each fit there
appears a trend where the spectrum is steeper below
15 keV than expected for a blackbody (see Fig.~2-4).
Combining the four strongest events together (Fig.~5) gives an unacceptable
$\chi^2$ (24.7 with 9 degrees of freedom).
This large $\chi^2$, due to the low energy steepness of the data,
is not caused by variations in the free
parameters from event to event since the results of variations in temperature
on the composite blackbody spectrum can only be to broaden it.
This trend of the blackbody producing the largest $\chi^2$ continues when
we add together the 5$^{th}$ to 14$^{th}$ and the 15$^{th}$ to the
40$^{th}$ events (Fig.~6, 7).
These large $\chi^2$ are not due to variations in the parameters
(which would
invalidate the combining of different events).
The other model fits show that the spectral shape remains remarkably constant.
Figure 1 also argues against a blackbody fit since
to
maintain a hardness ratio as flat as that of the linear fit to
figure 1, the emitting area must coincidently vary by a factor $>$ 15
while the temperature remains nearly constant.
We conclude that blackbody can be rejected as the explanation for the
low energy rollover seen in the PC data.

\subsec{4.2}{Thermal Bremsstrahlung - Blackbody}

Rejecting the blackbody shape does not necessarily mean that the rollover
is not due to self absorption.
To produce a blackbody spectrum, the  plasma must be optically thick
at all energies.
The plasma could be optically thick at low energy and thin at high
energies.
For free-free absorption with little electron scattering, the spectrum has
the form:
$$
\phi_{TB-BB}(E) = \phi_{BB}(E)\{1- e^{-(\phi_{TB}(E)/\phi_{BB}(E))}\}
\eqno(\new)
$$
Where, $\phi_{TB}$ is the optically thin thermal bremsstrahlung
continuum,
$$
\phi_{TB}(E) = 1.4\times 10^{-50} n_{e}^{2}l~\Psi
{}~E^{-1}(m_{e}c^{2}/kT)^{1/2}~e^{-E/kT} \eqno(\new)
$$
where $n_{e}$ is the electron density (cm$^{-3}$) and $l$ is the
thickness of the emitting region (cm).
This form gives an acceptable
fit to the data
with $kT  \sim 20$ keV and $\Psi = 13.8$ km$^2$ kpc$^{-2}$ for the strongest
events (see Table II).
However, the thermal bremsstrahlung process seems unlikely as the explanation
for the high energy spectrum since
$n_{e}^2 l$ is only 1.1$ \times 10^{51}$ cm$^{-5}$ and yet the plasma should
be optically
thin to Thompson scattering ($n_{e}l < 10^{24}$ cm$^{-2}$).
Although thermal bremsstrahlung has been used in fits
of SGRs as the continuum above 30 keV
(Mazets \etal 1982,
Atteia \etal 1987), it is only characteristic of the shape of the
continuum and is not necessarily the actual mechanism.
What the TB-BB spectral fit demonstrates is that the low energy rollover
could be a self-absorbed Rayleigh-Jeans spectrum even though blackbody
is unacceptable over the whole energy range.
The $1\sigma$ error bars on the temperature assume a single interesting
parameter.

\subsec{4.3}{Thermal Bremsstrahlung - Power Law}

To explore further what slopes below $\sim$ 15 keV are allowed we have fit
a power
law connected to thermal bremsstrahlung.
There were four free parameters: the slope of the power law ($\alpha$),
the connection energy ($E_c$), the thermal bremsstrahlung temperature (kT),
and an overall scale factor.
If $\alpha$ is 1 then this formulation could be considered a crude
Rayleigh-Jeans law with
a discontinuity at the peak where the formula abruptly switches from a
positive to a negative slope.
The slope below 15 keV is typically found to within $_-^+0.5$.
The best fit slopes vary from 0.2 to 1.5 with an average near 1.0.
In the fourth strongest event (Fig.~4) the $\chi^2$ surface had 2 local minima.
The curve labelled TB-PL had a $\chi^2$ of 10.0 and the one labelled TB-PL2
had a $\chi^2$ of 11.1; both should be considered acceptable.
The rather different functional forms that can fit the same data demonstrate
how obliging the data can be when the statistics become poor.
The disparity seen between TB-PL and TB-PL2 was a consideration
in our decision to analyze only
the first four events as individuals and to combine weaker events
together to obtain sufficient statistics.
The $1\sigma$ error bars on the power law index assume a single interesting
parameter.
Note from Table II that the derived thermal bremsstrahlung temperature
($\sim 22$
keV) is lower than earlier reports from {\it Prognoz 9} (40 keV, Atteia
\etal 1987)
and SMM (30 keV, Kouveliotou \etal 1987).

\subsec{4.4}{Thermal Bremsstrahlung - Photoelectric Absorption}

The rollover in a spectrum may be caused
by neutral photoelectric absorption which at these energies is
dominated by iron absorption.
Photoelectric absorption is a strong function of energy and can therefore
produce a rapid rollover.
We used Morrison and McCammon (1983) for our calculation of the absorption
which is based on cosmic abundances.
An distinctive feature of an absorption spectrum in this energy range is an
iron edge, but unfortunately, the sensitivity and energy resolution
in the PC channels is not enough to detect the iron edge with certainty.
Using a thermal bremsstrahlung continuum (Eq.~3) gives a
best fit value for $N_H$ of $1.1\times 10^{24}$
hydrogen atoms cm$^{-2}$ and kT $\sim$ 22 keV.
The normalization of the thermal bremsstrahlung gives $n_e^2l\Psi = 1.4
\times 10^{52}$ cm$^{-5}$ km$^2$ kpc$^{-2}$.
Both the temperature and column density are considered interesting
parameters so the $1\sigma$ error bars in Table II are based on changes
in $\chi^2$ of 2.3.
Within the statistics, all samples gave the same column density.
Such absorption is likely to be circumstellar and, therefore, rather constant.
Even a single example inconsistent with the derived column
density would be an argument against absorption being responsible
for the rollover. We have therefore checked individually all of the 20
strongest events and found no example that could not be fit by
$1.1 \times 10^{24}$ cm$^{-2}$.
This density is too high for molecular clouds.
It is possible that circumstellar material could have an
appropriate column density. GX301-2, for example, is in an accreting
binary that has a column density over $10^{24}$ cm$^{-2}$.  (White and Swank
1984).
Ejected surface material is another way to provide the matter
and is particularly tempting to consider since the source is
super Eddington.

The huge photon flux from the SGR threatens
to ionize a plasma near the source faster than it can recombine.
Assuming the thermal bremsstrahlung shape
used in the TB-PE fit (Fig.~2) extends
down to at least 5 keV, then $\sim 9 \times 10^{40}$ erg above 5 keV were
absorbed in the $1.1 \times 10^{24}$ H cm$^{-2}$ material.
If each ionizing photon is 8 keV, the number of iron-ionizing photons,
$n_{\gamma}$, is $\sim 7 \times 10^{48}$.
Events 831116-G and 831116-H occurred within one second and there is no
indication that the absorbing material was any different for the second
event.
The absorption is dominated by k-shell ionization of the iron which
makes up only a small fraction of the material.
The absorption would be turned off only if the iron is completely stripped
and if recombination is insufficient to replenish the k-shell electrons.
The total recombination coefficient for iron, $\alpha_{Fe}(kT)$, is
approximately $2 \times 10^{-11} (kT)^{-1/2}$ cm$^3$ sec$^{-1}$
where $kT$ is the temperature in keV of the electrons (Kaplan
and Pikelner, 1970).
The number of times that an iron atom can recombine per sec if it is
constantly being ionized is $n_{e} \alpha_{Fe}$ where $n_{e}$ is the electron
density.
To allow most of the iron to have filled k-shells within 1 second,
the number of ionizations and
recombinations per cm$^2$ must be
larger than the number of ionizing photons per cm$^2$, that is
$$
N_{Fe}\big(I_{Fe} + n_{e} \alpha_{Fe} \big) \gg {n_{\gamma} \over 4 \pi R_a^2}
$$
where $R_a$ is the radius of the absorbing material and $I_{Fe}$ is the
initial number of electrons per iron atom.
{}From the cosmic abundances
of Morrison and McCammon (1983), the column density of iron, $N_{Fe}$, is
$3.6 \times 10^{19}$ cm$^{-2}$.
(Note that the data effectively constrains $N_{Fe}$ rather than $N_H$.)~~
Using $n_{e} \sim N_H/R_a$ and $I_{Fe} \sim 20$,
one finds that $R_a$ must be at least a few
A.U.
If the material is iron rich, $N_{Fe}$ remains the same since it is the
amount of material necessary to provide the low energy rollover in SGR1806-20,
only $N_H$ (and $n_{e}$) is reduced.  As a result, $R_a$ is insensitive to the
ratio of iron to hydrogen in the plasma.

\subsec{4.5}{Other Fits}

Optically thin modifications to blackbody emission depend on the emission
process. Above we used free-free thermal bremsstrahlung.
By replacing $\phi_{TB}(E)$ in Eq.~2 with thermal synchrotron ($\phi_{TS}(E)$),
one obtains an optically thick/optically thin spectrum that is similar
to TB-BB.  Here
$$
\phi_{TS}(E) = 1.75\times 10^{-19} n_{e}l~\Psi
{}~(m_{e}c^{2}/kT)^{1/2}~e^{[-(4.5~E/E_s)^{1/3}]}  \eqno(\new)
$$
where $E_s = 11.6B_{12}(kT/m_{e}c^{2})^{2}$ and $B_{12}$ is the magnetic field
in units of $10^{12}$ Gauss ({\it cf} Liang, Jernigan, and Rodrigues 1983).
The curve labelled TS-BB in Fig.~8 is a best fit for self absorbed
thermal synchrotron.
The best fit $E_s$ is $\sim 0.20$ keV, comparable to the value found
for the March fifth event (0.30 keV, Liang 1986).
The Rayleigh-Jeans portion of $\phi_{BB}(E)$ is insensitive to the temperature
so the temperature is not found independently by fitting Eq.~2 to the data,
only $BT^2$ is constrained.
However, we do assume that the emission is well above the first cyclotron
harmonic so $B_{12}$ should be less than $\sim 0.5$.  Thus, the temperature
must be at least $\sim 30$ keV.
The temperature cannot be arbitrarily high since that would make the
magnetic field so low that it could not counteract the super Eddington
radiation pressure (see section 6.1).
Ignoring the problem of confining the plasma and only
addressing the assumptions behind Eq.~4, the temperature could easily
be as large as 100 keV.
The values of the other fit parameters depend on the temperature: $\Psi =
4.0 (60 {\rm keV}/kT)$ km$^2$ kpc$^{-2}$ and $n_el = 2.9 \times 10^{22}
(60 {\rm keV}/kT)^{-3/2}$ cm$^{-2}$.
The TS-BB fit gives an unacceptable $\chi^2$ (24 with 8 degrees of freedom)
but due almost entirely to the SC data: the data above 15 keV bends more than
allowed by Eq.~4.
Liang (1987) has proposed that injected electrons that cool could explain the
spectrum of SGR1806-20.  The cooling electrons tend to have more curvature
above 15 keV and, if the cyclotron fundamental is substantially below 15 keV,
self absorption could explain the rollover.  Liang compared such a model to
GB790107.  Above 15 keV there was general agreement
although below 15 keV there was insufficient statistics to detect the
rollover.  In a future paper we plan to fit such
models to the strong SGR events.

Mechanisms can produce a low energy rollover if
the emission process has a low energy cut off. For example,
thermal cyclotron is the sum of harmonically spaced, Doppler
broadened emission.
As such, below the first harmonic there is little emission, and the spectral
shape at low energies is
dominated by Doppler broadening.
The position of the peak is set by the magnetic field
and the emission below the peak would fall very rapidly if the temperature
parallel to the field is small.
Norris \etal (1991) argues in favor of such a process.
The relative emission at each harmonic can be calculated if one assumes an
electron distribution among the Landau levels ({\it e.g., }Brainerd
and Lamb, 1987).
In fact, the Brainerd and Lamb formulation gives an unacceptable $\chi^2$;
it predicts too much high energy emission.
However, it is unclear how any electron distribution is maintained when the
deexcitation timescale ($10^{-15}$ sec) is so much faster than the
collisional timescale.
Brainerd (1989) has suggested that radiative excitation might populate the
first excited state.
However, a single gaussian is inconsistent with the data;  it predicts too
little high
energy emission.  Presumably, radiative
excitation can populate several Landau levels (Norris \etal).
We have fit two harmonically spaced gaussians allowing the scale factor
between the first and second harmonic to be a free parameter
(see ``Gauss'' in Fig.~8).
That scale factor is related to the relative populations of the Landau levels.
The best fit magnetic field is $1.5 \times 10^{12}$ Gauss and the
Doppler widths are the order of 15 keV.
The fit is acceptable with a $\chi^2$ of 2.5 with 6 degrees of
freedom.
However, there are five free parameters so it is not too surprising that
there is an acceptable fit.

Various Comptonized spectra were tried.
A Wien peak (saturated Compton, $\phi(E)=E^2~e^{(-E/E_0)}$) fits poorly
yielding
a $\chi^2$ of 25.4 with 9 degrees of freedom.
Partially saturated Comptonization of a low energy source can be characterized
by the energy of the injected photons ($E_i$), the temperature of the
scattering medium ($kT_s$), and a parameter related to the optical depth.
Here we have implemented the formulation of Sunyaev and Titarchuk (1980).
The curve labelled ``Com-I'' in Fig.~8 used $E_i \sim 1$ keV and gave
8 keV for the best fit temperature of the scattering medium.
The Sunyaev and Titarchuk $\gamma$ parameter was 0.02.
The fit was unacceptable with $\chi^2$ = 35 for 8 degrees of freedom.
We also allowed the injected spectrum to be a blackbody with the temperature
a free parameter (see curve `Com-BB'' in Fig.~8).
The best fit parameters were $kT_s = 10$, $\gamma$ = 0.64, and the temperature
of the injected blackbody was 5.5 keV.
This fit gave an acceptable $\chi^2$ (10 with 7 degrees of freedom) although
there is a clear trend below the peak.
In general, Comptonized spectra fit poorly because the observations
imply a peak that is narrower than achievable with most Comptonization models.

The spectra of x-ray accreting pulsars are often represented as
$$
\phi_{Acc Pul}(E) = E^{\alpha-1}~~~~~~~~~~~~~~~~~~~~~~~~E < E_c
$$
$$
      ~~~~~~~~~~~~~~~~~= E^{\alpha-1}e^{(E_c-E)/E_F}~~~~~~~~E > E_c \eqno(\new)
$$
(see White, Swank, and Holt 1983).
This formulation is not directly derived from a physical model but gives
an adequate representation of the accreting pulsar spectrum.
The spectra of SGR1806-20 is similar to some accreting pulsars above
$\sim 15$ keV but
the accreting pulsars do not show a low energy rollover as strong as seen in
SGR1806-20.
Although SGR's and accreting sources may possibly occur near the
surface and involve cyclotron processes,
the super Eddington physics involved with an SGR is probably different
than that associated with objects that are just at the Eddington limit.

The cyclotron up-scattering process (CUSP,
Ho, Epstein, and Fenimore 1992, Dermer and Vitello, 1992)
produces spectra that have a low energy rollover at the temperature
of the underlying source of photons and a high energy rollover at $B^2/kT$.
If $B \sim kT$ then CUSP can produce a single peak.
Such a function fits poorly because the shape of the emission below the
peak has a Doppler half-width equal roughly to the temperature and it was
too broad to give an acceptable fit.

\sect{5. BURST INTENSITIES}

Table I summarizes the burst intensities for the four strongest events
defined by the sum of their counts in SC1 and SC2.
The fluxes are dependent on the durations of the events which are
available from Kouveliotou \etal (1987) and Atteia \etal (1987).
Previous work reported peak fluxes only for emission above 30 keV.
Our peak energy flux (for above 30 keV) is found by
integrating the best fit function for TB-PE above 30
keV (see Table II, Fig.~2).
The peak energy fluxes agree, on the average, with
those of Kouveliotou \etal and Atteia \etal to within
10\%.
The spectra fall off rapidly both below 15 keV and above 50 keV.
Thus, the total flux can be estimated by integrating the best
fit TB-PE function
over all energies ({\it e.g., }5.3$\times 10^{-5}$ erg cm$^{-2}$
sec$^{-1}$ for the brightest event, Table I).
About 2/3 of the total flux comes from below 30 keV.
Kulkarni and Frail (1993) suggested that G10.0-0.3, a supernova remnant
at 17 kpc, is associated with SGR1806-20
and Murakami \etal (1994) has observed a
SGR1806-20 event centered on the remnant.
Using this distance one can estimate the total luminosity of the events.
The brightest ICE events had a luminosity of $1.8 \times 10^{42}$
erg sec$^{-1}$, about $2\times 10^{4}$ times the
Eddington limit.
Although that is extremely luminous, the March fifth event was a factor of 450
times more luminous (above 30 keV).
In the bandpass of 5 to 50 keV, the brightest SGR1806-20 event
rose to 300 Crab and turned off again within $\sim$ 0.1 sec.

\sect{6.~DISCUSSION}

\subsec{6.1}{Super Eddington Fluxes in a Magnetic Field}

The identification of SGR1806-20 with the supernova remnant G10.0-0.3
has
given us a distance
($\sim 17$ kpc) and therefore a total luminosity which peaks
at $\sim2 \times 10^{42}$ erg sec$^{-1}$ (see Table I).
The Eddington Limit, $L_E$, is the maximum luminosity where the radiation
pressure does not exceed the gravitation force.  Simplistically,
$$
L_E \sim {4\pi G M m_H c \over \sigma_{Th}}  \sim 10^{38}
{\rm ~~erg ~~sec^{-1}}  \eqno(\new)
$$
where $G$ is the gravitational constant, $M$ is the mass of the star,
$m_H$ is the mass of a proton,
$c$ is the speed of light,
and $\sigma_{Th}$ is the Thompson cross section.
 SGR1806-20 has produced
events that are $\sim 2 \times 10^4$ times the Eddington Limit.
The SGR super Eddington fluxes can last from 100 ms to 200 sec,
much longer than
the dynamic time scale of a neutron star.
Such a super Eddington flux makes accretion models unlikely (although
not all authors would agree, see {\it e.g.,} Colgate and Leonard, 1993).
The initial angular momentum of an accreting body and tidal forces would
likely break up a body such that it is susceptible to radiation pressure.
Some explanations have
depended on the magnetic field to either confine the plasma and/or reduce
the opacity such that the radiation cannot blow the accreting material
away.  The magnetic field falls off rapidly ($R^{-3}$)
but the Eddington Limit is independent of radius so
it seems
unlikely that magnetic field effects will allow accretion with typical
impact parameters to be sustained for many dynamic timescales in the presence
of a super Eddington flux.

We favor an internal (yet unspecified) source of energy.
The super Eddington flux is still a concern for internal sources since the
pressure should blow away the energy producing material.
However, at the surface the magnetic field can modify the Eddington limit
in two ways.  The field might be able to provide a geometery-dependent
confining pressure
(Lamb 1982) or
a super strong
magnetic field ($\sim 10^{14}$ Gauss) can reduce the opacity at low
energy where the SGRs radiate (Paczy{\'n}ski 1992).

First we consider the effect of the magnetic field on the opacity.
At energies much less than the cyclotron fundamental, $E_{cyc}$, the
Compton cross section follows
$$
\sigma_{cyc,1} \approx \sigma_{Th}\bigg(\sin^2\theta +\cos^2\theta
(E/E_{cyc})^2\bigg)
$$
$$
\sigma_{cyc,2} \approx \sigma_{Th} (E/E_{cyc})^2    \eqno(\new)
$$
where $\sigma_{cyc,1}$ and $\sigma_{cyc,2}$ are for the two linear
polarization states
and $\theta$ is the angle of the
photon relative to the magnetic field (Herold 1979).
Paczy{\'n}ski (1992) suggested that the Rosseland mean opacity be used
in Eq.~6.
 The Rosseland mean gives the most weight to the lowest
opacity and could reflect how the radiation evolves to escape along the opacity
windows of least resistance.
For unpolarized light the lowest opacity is along the field lines
and it is
unlikely that the geometry is such that we are always looking along the
field lines.
(We see many bursts from 1806-20 and sustained pulsations from the
SGR phase of the March 5 event.)
The flux-weighted mean should be used in Eq.~6 (see {\it e.g.}, Mihalas
1970).
The Eddington Limit in the
presence of a super strong magnetic
field, $L_{E,B}$, is
$$
L_{E,B} = L_E {\int_0^{\infty}\sigma_{Th}E\phi(E)dE \over \int_0^{\infty}
\sigma_{cyc} E\phi(E)dE }   \eqno(\new)
$$
if $\phi(E)$ is effectively zero for $E~~\gax~~E_{cyc}$.
Note that, here, the observed spectrum is used ({\it cf.}  Eqs.~1-5).
Let $\psi_1$ be the fraction of the radiation that is in linear
polarization state 1, and $\psi_2 = 1-\psi_1$ then
$$
{L_{E,B} \over L_E}  = {
\int_0^{\infty}E\phi(E)dE  \over
\psi_1\int_0^{\infty} \big(\sin^2\theta +\cos^2\theta
({E \over E_{cyc}})^2 \big) E\phi(E)dE +
\psi_2\int_0^{\infty} ({E \over E_{cyc}})^2 E\phi(E)dE     }
       \eqno(\new)
$$
For unpolarized light, the Rosseland mean gives a large $L_{E,B}$ whereas
the flux-weighted mean gives $L_{E,B} \sim L_E(\psi_1\sin^{2}\theta)^{-1}$
where $\psi_1\sin^{2}\theta$ is the order of unity.
Thus, we suggest that in order to have $L_{E,B} \sim 10^4 L_E$ (as required
by the observations), $\psi_1$ must be the order of $10^{-4}$; the radiation
must be completely polarized.
Perhaps the polarization state evolves until all of the radiation is
in the polarization state that can escape.

Our observations could completely define $\phi(E)$ since it falls off
rapidly at both high and low energy.  Significant unmeasured emission below our
lowest observation ($\sim 5$ keV) would not affect Eq.~9.  Unmeasured
emission above our highest observation ($\sim 1.2$ MeV) is unlikely.
We can evaluate Eq.~9 for the various spectral shapes that give
acceptable fits to the data and determine the minimum $E_{cyc}$ that gives
$L_{E,B} = 2 \times 10^4 L_E$.  The resulting $E_{cyc}$'s vary from 4700
keV for
the two gaussian fit to 5700 keV for the BB-TB fit. The corresponding
magnetic fields
are the order of $4-5 \times 10^{14}$ Gauss.  This is the same value as
obtained by
Paczy{\'n}ski except we require that the radiation be completely polarized
and we have some confidence that $\phi(E)$ is completely known.

Alternatively, the magnetic field can exert a pressure which assists the
gravitational forces in retaining the plasma near the surface.
The confining forces depend on the geometry and it is not clear if the plasma
can be confined on open field lines such as might be found near the poles.
Lamb (1982) suggested that the field necessary to confine the plasma can
be found from
$$
\beta {4 \sigma T^4 \over c} \ll {B^2 \over 8\pi}   \eqno(\new)
$$
where $\sigma$ is the Stefan-Boltzmann constant and $\beta$ depends on
the angular distribution of the radiation field; it is 1/3 for isotropic
radiation.
Here, we have required that the magnetic field energy density be much larger
than the radiation pressure since the magnetic field does not confine the
plasma along the field lines.
{}From Eq.~10, the requisite magnetic field can be estimated
$$
B_{12} \gg \bigg[{T \over 170}\bigg]^2   \eqno(\new)
$$
where T is in keV.
For example, Norris \etal used 30 keV
implying a
field of $B_{12} \sim 0.03$.
This estimate is valid only if the process is blackbody.
A lower limit on the requisite field can be found from
the radiation pressure on the outer photosphere, that is, the region where
the observed spectrum is
formed ({\it cf.} Fenimore, Klebesadel, and Laros 1984):
$$
B^2 \gg \beta {32 \pi \over c} \bigg({D \over R}\bigg)^2
\int_0^{\infty}E\phi(E)dE  .  \eqno(\new)
$$
Here, $D$ is the distance and $R$ is the radius of the source.
Substituting standard values gives
$$
B_{12} \gg  (D/{\rm kpc})(R/{\rm km})^{-1} I^{1/2}_{total}  .\eqno(\new)
$$
This expression is very useful since it depends only on the observed
intensity and does not make the assumption that the process is blackbody.
{}From Table I, $I_{total}$ is $5.3 \times 10^{-5}$ erg sec$^{-1}$ cm$^{-2}$
and using $D = 17$ and
$R = 10$, gives $B_{12} = 0.012$.

In comparing estimates of the magnetic field from Eqs.~11 and 13, one should
note that Eq.~11 uses the observations to determine the temperature and
does not require knowledge of the distance or the radius of the source.
However, Eq.~11 assumes that the shape of the spectrum (and the process
generating the photons) is equivalent to a blackbody.
Our fits involved "temperatures" that varied by an order of magnitude from
9 keV for a blackbody to $\sim$ 100 keV for self absorbed thermal synchrotron
requiring fields of $B_{12}$ = 0.003 to 0.3.
 Equation 13 uses the observations to determine
the shape of the spectrum and
is useful if
some estimate of the distance and radius is possible.
Eq.~13 only limits the outer photosphere and deeper in
the photosphere the spectrum
might be closer to blackbody with the temperature that
characterizes the optically thin observed emission making Eq.~11
more appropriate. This is uncertain since we do not
know the process that generates the observed spectrum nor if it ever takes
on the characteristics of a blackbody deeper in the photosphere.
For example, there could be an electron acceleration process that
produces an maxwellian velocity distribution with a particular temperature but
the radiation field may
not come into equilibrium with it.
In that case, Eq.~13 would be more appropriate to use.
Equation 13 is a lower limit and in most cases (SGRs, x-ray bursts, classic
gamma-ray bursts) we have fairly good estimates of $I_{total}$ although
not always $D$ and $R$. Equation 11 requires the temperature and that is model
dependent.
In either
case, whether the field can truly confine the plasma through magnetic pressure
depends on the geometry of the field lines.

\subsec{6.2}{The Role of the Magnetic Field in the Emission Process}

There is general agreement that the magnetic field plays a crucial role
in the SGR phenomenon
(Laros \etal 1986, 1987, Liang 1987, Norris \etal 1991, Paczy{\'n}ski 1992,
Colgate and Leonard 1993).
The range of fields that have been suggested spans nearly a factor of
$\sim 10^4$.
Here we discuss these suggestions in the context of our spectral observations.

A very strong field ($\sim 5 \times 10^{14}$ Gauss) suppresses
the opacity and allows a
super Eddington flux (Paczy{\'n}ski 1992).  If that is the case,
then the two-gaussian
fits and the self-absorbed synchrotron fits are irrelevant
since they require much smaller fields.
The remaining potential process
that can produce the continuum is thermal bremsstrahlung.
Self-absorbed thermal bremsstrahlung requires a large
photosphere ($\Psi \sim
30$ km$^2$kpc$^{-2}$, radius $\sim 50$ km, see Table I)
which would require a surface field of $\sim 6 \times 10^{16}$ Gauss, probably
an unreasonable value for a 10 km radius surface.
Thus, the large field assumption might be one way to explain the macroscopic
issue of the super Eddington flux but the process generating the photons
is still unclear.
Whatever the  process is, it must be capable of producing 100\% polarized
radiation, take place near the surface, and operate well below the cyclotron
fundamental.
One possible process is to consider that the electrons are always in
the ground Landau state such that they act as beads on strings.
The relative motions follow an one-dimensional maxwellian and the electrons
radiate in a manner similar to free-free bremsstrahlung.

The two gaussian fit could be evidence that the process is
low harmonic emission
from
a magnetic field of $\sim 2 \times 10^{12}$ Gauss (Norris, \etal 1991).
The low energy cutoff is then related to the Doppler width of the first
harmonic, that is, the parallel temperature of the emitting electrons.
It is unclear how the excited Landau levels remain populated.
A $2 \times 10^{12}$ Gauss field would not reduce the opacity sufficiently
to allow the super Eddington flux so the super Eddington flux must be
counteracted by magnetic field pressure.

A low field ($\sim 10^{11}$ Gauss) might be able to explain
the observed continuum through self absorbed synchrotron radiation
(Liang 1986, Liang 1987).
Either thermal or injected electrons radiate in high harmonics but merge
into the Rayleigh-Jeans continuum at $\sim 15$ keV.
The surface area ($\Psi \sim 4$ km$^2$ kpc$^{-2}$, radius $\sim 10$ km)
is reasonable.
The assumption that the first harmonic is at a lower energy than where
the emission is self absorbed requires the magnetic field to be less than
$\sim 5 \times 10^{11}$ Gauss. Equation 11 (which
limits $BT^{-2}$) combined with the $E_s$
value of 0.2 keV (which
limits $BT^2$) requires the field to be much larger than $1.2 \times
10^{11}$ Gauss.  Thus, the field required for self absorbed thermal
synchotron might be too small to confine the plasma against the
super Eddington flux.  Only the lower limit from Eq.~13
($1.2 \times 10^{10}$ Gauss)
is consistent with both confining the plasma and generating the spectrum.

\subsec{6.3}{The Nature of the Low-Energy Rollover}

We have presented three different explanations for the low energy rollover
seen in SGR1806-20: the Doppler broadened emission from cyclotron line
emission, self-absorption by either thermal or synchrotron processes, or
absorption by circumstellar material.  The two-gaussian fit is the best
evidence for the Doppler broadened emission explanation and more work will
be necessary to develop a complete model of this process.  It might have the
advantage that it could explain the fact that the spectrum seems identical for
events that vary in intensity by a factor of 50.  The spectra would be very
similar if the ratio of emission in the second harmonic compared to the
first is set by relative probabilities of transitions involving two Landau
levels to that involving one.

Photoelectric absorption by neutral
material requires
a column density of $10^{24}$ Hydrogen atoms cm$^{-2}$.
Such a high column density is rarely found except in the cores of extremely
dense molecular clouds.
Figure 5 of Sanders \etal (1986) does not show dense molecular clouds in the
direction of SGR1806-20 so we conclude that, if photoelectric absorption
is involved, it is due to circumstellar material.
This material cannot be close to the origin of the bursts since the iron
would be completely ionized by the radiation (see section 4.4).
The material must be a few A.U. away from the source.
Assuming spherical symmetry, the amount of material is
quite large:
$\sim 4 \pi R_a^2 \zeta N_H$ or
$\sim 2 \times 10^{-6} \zeta M_{\odot} (R/{\rm A.U.})^2$
where $\zeta$ is the ratio of the amount of iron in the plasma to what
one would expect with a cosmic abundance.
If the circumstellar material extends closer to the central source,
then weak bursts would ionize it less than strong bursts resulting in a
correlation between intensity and low energy emission: weaker bursts would
have less low energy emission. Although the uncertainties are large,
Fig.~7 tends to show the opposite, that the emission below the peak
tends to be more for weaker bursts.
Thus, there is no reason to suspect that the absorbing material is close
to the central source thereby reducing the amount of material necessary:
$R_a$ could be larger than a few A.U.~requiring between
$10^{-5}$ and  $10^{-4} \zeta M_{\odot}$
of material.
However, $\zeta$ could be quite small if the material is very iron rich.

Murakami \etal (1994) has suggested that a steady source observed by Asca
(AX1805.7-2025) is also the origin of the SGR1806-20 events.
The steady source has a spectrum that is absorbed by neutral material
but with a column density of only $10^{22} N_H$ cm$^{-2}$.
If the steady source is also the SGR source then the rollover
is probably not due to absorption by neutral material since it is unlikely
that the column density due to material a few A.U.~
from the source would change by two orders of magnitude.
Independent of the cause of the low-energy roll over we report in this paper,
it appears likely that
the steady source seen by Asca is not the SGR1806-20 source but is
the plerion itself acting as a foreground object. Then the $10^{22} N_H$
cm$^{-2}$
reported by Murakami \etal
is best explained as the typical interstellar absorption in this direction.
The actual SGR source would be buried deeper in the plerion and have
a substantial amount of circumstellar material causing the low energy rollover
we observe.
In fact, if the roll over is due to $10^{24} N_H$ cm$^2$, then any typical
steady
emission from the neutron star would be hidden from most instrumentation.
The strongest argument against the low energy rollover being due to neutral
absorption is the rather large amount of material that would be required.

Self absorption could arise from either thermal or synchrotron processes.
A completely optically thick spectrum ({\it i.e.,} a blackbody spectrum)
does not seem to give an acceptable fit (Table II).
However, an optically thin/optically thick
thermal bremsstrahlung spectrum assumes some self-absorption
and can give an acceptable fit (see Table II).
Although optically thin/optically thick synchrotron did not fit as well
(see section 4.5), more detailed calculations involving injected electrons
or cooling distributions ({\it cf.} Liang 1987) might be able to fit.
For free-free self absorption, $n_e^2l$ is $1.1 \times 10^{51}$ cm$^{-5}$.
The assumption of being optically thin implies
that $\sigma_{Th}n_el$ is less than unity so $l \sim 10^{-3}$.
The very thin regions required by free-free emission is an argument
against thermal bremsstrahlung as the process producing the photons.
For self absorbed thermal
synchotron, $n_el$ is $2.9 \times 10^{22}$ so it is self consistent with
the assumption of being optically thin and there are no constraints on
the thickness of the emitting region.

The spectrum above the peak is remarkably similar implying a constant
temperature yet the intensity varies by a factor of 50.  This
implies that the area is the only parameter that changes from burst to burst.
This is very unlike other transient events thought to occur on neutron stars
(such as Type I x-ray bursts) that usually have a somewhat constant
temperature but varying emitting area.
Using the distance to G10.0-0.3 and the $\Psi$ parameter for
the brightest events,
the maximum size of the emitting region has a radius
of $\sim 50$ km for the TB-BB fit and  $\sim 10$ km for the TS-BB fit, a more
reasonable value for a
neutron star.
The emission is probably near the surface for two reasons: (1)~it was near
the surface during the SGR phase of the March 5th event (the pulsations)
and (2)~if one
uses the magnetic field to confine the plasma against the super Eddington
flux, then the surface is the best place to do that since the magnetic field
falls off as $R^{-3}$ whereas the Eddington flux is independent of distance.

{\it Acknowledgement} This work was done under the Auspices of the US
Department of Energy.
We thank R.~Epstein, C.~Ho, C.~Kouveliotou, E.~Liang,
B.~Paczy{\'n}ski, and J.~Norris
for extremely helpful discussion and important corrections to the manuscript.

\vfill\eject

\sect{REFERENCES}

\par\noindent\hangindent=3pc\hangafter=1
Anderson, K. A., \etal 1978, IEEE Trans. Geosci. Electron., { GE-16}, 157.

\par\noindent\hangindent=3pc\hangafter=1
Atteia, J. L., \etal 1987 Ap. J. Lett. { 320}, L105-L110.

\par\noindent\hangindent=3pc\hangafter=1
Band, D., \etal 1993 Ap. J. { 413}, 281-292.

\par\noindent\hangindent=3pc\hangafter=1
Brainerd, J. J., 1989 Ap. J., Lett. { 341}, L67-L69.

\par\noindent\hangindent=3pc\hangafter=1
Brainerd, J. J. and Lamb, D. Q., 1987 Ap. J., { 313}, 231-262.

\par\noindent\hangindent=3pc\hangafter=1
Cline,T. L., \etal 1982, Ap. J. Lett., { 255}, L45-L48.

\par\noindent\hangindent=3pc\hangafter=1
Colgate, S.~A., and Leonard, P.~J.~T. 1993
{\it Proc. of Second Hunstville Conference on Gamma Ray Bursts}, in press.

\par\noindent\hangindent=3pc\hangafter=1
Dermer, C. and Vitello, P. 1992 in
{\it Gamma-Ray Bursts Observations, Analyses
and Theories} eds C. Ho, R. I. Epstein, and E. E. Fenimore (Cambridge:
Cambridge University Press) p 321-328.

\par\noindent\hangindent=3pc\hangafter=1
Evans, W. D., \etal 1980, Ap. J. (Letters), { 237}, L7-L9.

\par\noindent\hangindent=3pc\hangafter=1
Fenimore, E.E., Evans, W. D., Klebesadel, R. W., Laros, J. G., and
Terrell, J.  1981, Nature { 289}, 42-43.

\par\noindent\hangindent=3pc\hangafter=1
Fenimore, E.~E., Klebesadel, R. W., and Laros, J. G. 1983,
Adv. Space Res. Vol 3.,
No.4, 207-210.

\par\noindent\hangindent=3pc\hangafter=1
Fenimore, E.~E., Klebesadel, R. W., and Laros, J. G. 1984, in
{\it High Energy Transients in Astrophysics} ed. S. E. Woosley, AIP Conf.
115 (New York: American Institute of Physics) p 590-596.

\par\noindent\hangindent=3pc\hangafter=1
Fishman, G. \etal, 1993, Ap. J. Suppl. Series accepted.

\par\noindent\hangindent=3pc\hangafter=1
Golenetskii, S. V., Iiyinskii, V. N., and Mazets, E. P.  1979,
Nature { 307}, 41-43.

\par\noindent\hangindent=3pc\hangafter=1
Herold, H. 1979,
Phys.~Rev.~D  { 19} 2868-2875.

\par\noindent\hangindent=3pc\hangafter=1
Ho, C., Epstein, R. I., and Fenimore, E. E. 1992 in
{\it Gamma-Ray Bursts Observations, Analyses
and Theories} eds C. Ho, R. I. Epstein, and E. E. Fenimore (Cambridge:
Cambridge University Press) p 197-304.

\par\noindent\hangindent=3pc\hangafter=1
Kaplan, S. A, and Pikelner, S. B. 1970, {\it The Interstellar Medium}
(Cambridge Mass. Harvard University Press)
p 413.

\par\noindent\hangindent=3pc\hangafter=1
Klebesadel, R. W. 1992, in
{\it Gamma-Ray Bursts Observations, Analyses
and Theories} eds C. Ho, R. I. Epstein, and E. E. Fenimore (Cambridge:
Cambridge University Press) p 161-168.

\par\noindent\hangindent=3pc\hangafter=1
Kouveliotou, C., \etal 1987 Ap. J. (Letters) { 322}, L21-L25.

\par\noindent\hangindent=3pc\hangafter=1
Kouveliotou, C., \etal 1993a Nature 362 728-730.

\par\noindent\hangindent=3pc\hangafter=1
Kouveliotou, C., \etal 1993b Ap. J. 413 L101-L104.

\par\noindent\hangindent=3pc\hangafter=1
Kouveliotou, C., \etal 1994 Nature, submitted.

\par\noindent\hangindent=3pc\hangafter=1
Kulkarni, S. R., and Frail, D. A. 1993 Nature { 365} 33-35.

\par\noindent\hangindent=3pc\hangafter=1
Lamb, D.~Q. 1982, in
{\it Gamma Ray Transients and Related Astrophysical Phenomena}
eds. R. E. Lingenfelter, H. S. Hudson, and D. M. Worrall
AIP Conf. 77 (New York:
American Institute of Physics) p 249-272.

\par\noindent\hangindent=3pc\hangafter=1
Lampton, M., Margon, B., and Bowyer, S. 1976 Ap. J. { 208} 177-190.

\par\noindent\hangindent=3pc\hangafter=1
Laros, J. G., \etal 1986, Nature, { 322}, 152-153.

\par\noindent\hangindent=3pc\hangafter=1
Laros, J. G., \etal 1987, Ap. J. (Letters), { 320}, L111-L115.

\par\noindent\hangindent=3pc\hangafter=1
Laros, J. G., \etal 1990, XXI Internat. Cosmic. Ray Conf., { 1}, 68.

\par\noindent\hangindent=3pc\hangafter=1
Lewin, W. H. G., and Joss, P. C. 1983 in {\it Accretion Driven Stellar
X-ray Sources} eds. W. H. G. Lewin and E. P. J. van den Heuvel
(Cambridge: Cambridge University Press), p 41.

\par\noindent\hangindent=3pc\hangafter=1
Liang, E. P., Jernigan, T. E., and Rodrigues, R. 1983 Ap. J. { 271}, 766-777.

\par\noindent\hangindent=3pc\hangafter=1
Liang, E. P. 1986 Ap. J. Lett. { 308}, L17-L20.

\par\noindent\hangindent=3pc\hangafter=1
Liang, E. 1987, in
{\it Nuclear Spectroscopy of Astrophysical Sources}
AIP Conf. 170 (New York:
American Institute of Physics) p 302-306.

\par\noindent\hangindent=3pc\hangafter=1
Loredo, T., and Epstein, R. I., Ap. J., { 336}, 896-919.

\par\noindent\hangindent=3pc\hangafter=1
Mazets, E. P., Golenetskii,
 S. V., Iiyinskii,
 V. N., Aptekar, R. L., and Guryan, Y. A., 1979, Nature, { 282} 587-589.

\par\noindent\hangindent=3pc\hangafter=1
Mazets, E. P., Goleneskii, S. V., Guryan, Y. A., 1979
Sov. Astron Let, 5(6), Nov-Dec,  343-344.

\par\noindent\hangindent=3pc\hangafter=1
Mazets, E. P. and Golenetskii, S. V., 1981 Astrophys. Space Sci. { 75} 47-81.

\par\noindent\hangindent=3pc\hangafter=1
Mazets, E. P., Goleneskii, S. V., Guryan, Y. A., Il'inskii, V. N. 1982
Astrophys. Space Sci., { 84}, 173-189.

\par\noindent\hangindent=3pc\hangafter=1
Mihalas, D.  1970, {\it Stellar Atmospheres}
(San Francisco. Freeman and Company)
p 38.

\par\noindent\hangindent=3pc\hangafter=1
Morrison, R. and McCammon, D. 1983,Ap. J., { 270}, 119-122.

\par\noindent\hangindent=3pc\hangafter=1
Murakami \etal, Y. 1994 Nature, submitted.

\par\noindent\hangindent=3pc\hangafter=1
Norris, J. P., Hertz, P., Wood, K. S., and Kouveliotou, C. 1991,
Ap. J., { 366}, 240-252.

\par\noindent\hangindent=3pc\hangafter=1
Paczy{\'n}ski, B. 1992, Acta Astronomica { 42} 145-153.

\par\noindent\hangindent=3pc\hangafter=1
Petrosian, V. 1981, Ap. J., { 251}, 727 738.

\par\noindent\hangindent=3pc\hangafter=1
Sanders, D. B., Clemens, D. P., Scoville, N. Z., and Solomon, P. M. 1986
Ap. J. Supp. { 60} 1-296.

\par\noindent\hangindent=3pc\hangafter=1
Sunyaev, R. A. and Titarchuk, L. G. 1980 Astron. Astrophys. { 86} 121-138.

\par\noindent\hangindent=3pc\hangafter=1
Ulmer, A., Fenimore, E. E., Epstein, R. I., Ho, C., Klebesadel, R. W.,
Laros, J. G., and Delgato, F. 1993, Ap. J. 418 395-397.

\par\noindent\hangindent=3pc\hangafter=1
White, N. E., Swank, J. H., and Holt, S. S. 1983, Ap. J., { 270}, 711-734.

\par\noindent\hangindent=3pc\hangafter=1
White, N. E.,  and Swank, J. H. 1984, Ap. J., { 287}, 856-867.
\vfill\eject

\sect{FIGURE CAPTIONS}

{\bf Fig.~1}: Hardness Ratios for 95 SGR1806-20 events seen by ICE as a
function of fluence.
Hardness is the ratio of the 43.2 to 77.5 keV energy channel to the
25.9 to 43.2 channel. The solid line is a best fit linear function.
There is very little variation over a dynamic range of 50
implying that only the emitting area is changing from burst to burst.

{\bf Fig.~2}: Different spectral fits to the sum of the largest and third
largest SGR1806-20 event.  BB is blackbody, TB-BB is optically
thin thermal bremsstrahlung merged into a blackbody,
TP-PL is
thermal bremsstrahlung connected to a power law,
 and TB-PE is
thermal bremsstrahlung with photoelectric absorption by $\sim 10^{24}$
Hydrogen atoms cm$^{-2}$ with a cosmic abundance.

{\bf Fig.~3}: Same as Fig.~2 except for the second strongest SGR1806-20
event.

{\bf Fig.~4}: Same as Fig.~2 except for the fourth strongest SGR1806-20
event.

{\bf Fig.~5}: Same as Fig.~2 except for the sum of the first four
strongest SGR1806-20 events.

{\bf Fig.~6}: Same as Fig.~2 except for the sum of the 5$^{th}$ to the
14$^{th}$ strongest SGR1806-20 events.

{\bf Fig.~7}: Same as Fig.~2 except for the sum of the 15$^{th}$ to the
40$^{th}$ brightest SGR1806-20 events.

{\bf Fig.~8}: Spectral fits to the largest and third largest SGR1806-20
events.
TS-BB is an optically thick/optically thin spectrum based on thermal
synchrotron, Gauss is the sum of two Gaussians, Wien is a saturated Comptonized
spectrum, Com-In is Comptonization of soft injected photons, Com-BB
is a Comptonized blackbody, and
ACC PUL is an accreting pulsar spectrum.
\vfill\eject
\bye